\documentclass[12pt, letterpaper]{article}
\usepackage{booktabs} 
\usepackage[ruled]{algorithm2e} 

\usepackage[margin=1in]{geometry}
\usepackage{setspace}
\doublespace     
\usepackage{natbib}
\usepackage{amsmath}
\usepackage{graphicx}
\usepackage{authblk}

\usepackage{amsthm}
\newtheorem{theorem}{Theorem}
\newtheorem{lemma}[theorem]{Lemma}
\newtheorem{proposition}[theorem]{Proposition}

\title{Incentives for Digital Twins: Task-Based Productivity Enhancements with Generative AI}
\author[1]{Catherine Wu}
\author[1]{Arun Sundararajan}
\affil[1]{Information Systems, NYU Stern}

\date{May 20, 2025}

\begin{document}
\maketitle

\begin{abstract}
Generative AI is a technology which depends in part on participation by humans in training and improving the automation potential. We focus on its ability to replicate the process or style of an \textit{individual} human worker. This kind of ``AI twin'' could complement its creator’s efforts, enabling them to produce higher-quality output in their individual style more efficiently. However, increasingly intelligent AI twins could also, over time, replace individual humans. We analyze this trade-off using a principal-agent model in which agents have the opportunity to make investments into training an AI twin that lead to a lower cost of agent effort, a higher probability of success, or both. We propose a new framework to situate the model in which the tasks performed vary in the ease to which AI output can be altered or improved by the human (the “editability” of the output) and also vary in the extent to which a non-expert can assess the quality of human or AI output (its “verifiability.”) Our synthesis of recent empirical studies indicates that productivity gains from the use of generative AI are higher overall when task editability is higher, while non-experts enjoy greater relative productivity gains for tasks with higher verifiability. 

\medskip
We show that during investment decisions, paradoxically, a strategic agent will trade off improvements in output quality and ease of effort to preserve their wage bargaining power. Tasks with high verifiability and low editability are strategically most aligned with a worker’s incentives to “train their twin,” but for tasks where the stakes are low, this alignment is constrained by the agent holding back to mitigate their risk of displacement. We extend these results in a two-period model and show that settings in which training investments yield the highest productivity gains also put the non-strategic agent at the greatest risk of mistakenly creating an AI twin that will displace them in the long run. Our results suggest that sustained improvements in company-sponsored generative AI will require nuanced design of human incentives, and that public policy which encourages balancing worker returns with generative AI improvements could yield more sustained long-run productivity gains. 
\end{abstract}

\section{Introduction}

In the early days of computer-based factory automation, a human machine operator's manual actions were recorded, often via punched cards or tapes, and then "played back" to drive machine tools automatically. These "teach-and-repeat" methods laid the foundations for the computer numerical control (CNC) systems that powered an early wave of computer-driven automation \citep{_1991_CNC}. Today, although generative AI is largely viewed as general-purpose source of knowledge and skills, a distinguishing characteristic of the technology is its ability to replicate the process or style of an \textit{individual human creator}. A human's body of work can now serve as training data to create AI that emulates their unique approaches, skills, and creative processes. From replicating an author's writing style to capturing a surgeon's precise movements, these "digital twins" \citep{Heffernan_2024_ArtificialIntelligenceHuman} generate the ability to scale human expertise in unprecedented ways, while simultaneously raising new challenges to an individual's ownership of their human capital. The potential of a new wave of replication-driven displacement extends beyond creative domains and into highly skilled professional fields, where complex individual human capabilities may be systematically captured and partially reproduced by AI systems, leading to a potential reversal of the skill-biased technological change that has characterized digital progress over the last decades \citep{Acemoglu_2002_TechnicalChangeInequality, Acemoglu_2021_HarmsAI}. 

Platforms like Wizly\footnote{https://askwizly.ai/} that allow individuals to create ``AI twins'' and use them to scale their independent work will no doubt proliferate in the coming years. Thus, training an AI system to replicate individual skills may augment human potential, improving labor market outcomes and reducing inequality \citep{Agrawal_2023_TuringTransformationArtificial}. A custom-trained generative AI model could augment its creator's skill in their own domain, enabling the creator to produce higher-quality output faster and more efficiently. At the same time, workers may view increasingly intelligent tooling as a threat that could replace them or lower their ability to command premium wages. 

How will the net effect of these opposing forces affect the incentives for a knowledge worker to participate in the creation of their AI twin? This remains an open question, one that depends on factors that include the relative rate of quality improvements an AI twin provides with or without the human in the loop, the impact of individually customized AI systems on a human's cost of effort, and the benefits that an organization enjoys from outcomes of varying levels of quality and success. 

While there is a growing body of literature about whether artificial intelligence will substitute for or complement human labor across various sectors \citep{Noy_2023_ExperimentalEvidenceProductivity, Felten_2023_OccupationalHeterogeneityExposure, Brynjolfsson_2023_GenerativeAIWork}, research into the incentives a human worker might have to participate in the creation and improvement of such systems is limited. Moreover, recent projections about the impact of AI on worker productivity are mixed. A study conducted by The Upwork Research Institute reports that “77\% [of employees] say [AI] tools have actually decreased their productivity and added to their workload” \citep{temporary-citekey-177}. In contrast, a study by McKinsey and Company estimate that generative AI “could add the equivalent of \$2.6 trillion to \$4.4 trillion annually” to the economy \citep{temporary-citekey-185}.  Because AI is a general purpose technology \citep{Eloundou_2023_GPTsAreGPTs}, it is valuable to assess how these shifts of productivity will manifest in the long-run in market structure and governance \citep{Tucker_2024_HowDoesCompetition}.

In this paper, we model a human worker who can partner with an AI system to accomplish tasks assigned by a manager. The capabilities of the AI system are inferior to those of the human worker, but the worker can choose to invest in training it to be an AI twin in order to either reduce the effort they need to exert, to improve work outcomes when partnering with the AI system, or both. [A more detailed paper with mathematical details and proofs is available on request.]

We first consider a short-run problem, investigating how the worker's level of investment in the creation of their AI twin depends on intrinsic task properties---specifically, (i) how easily a human worker can modify AI output to reach a desired quality level, which we term the \textit{editability} of the task, and (ii) whether the manager is able to verify the quality of AI output absent the human worker, which we term the task's \textit{verifiability}. We define these properties based on their ability to explain recent findings from the empirical literature connecting the use of generative AI to productivity. We situate our problem in a two-outcome limited liability moral hazard model with hidden action. The manager (principal) chooses whether or not to offer the capability to customize an AI system to the worker (agent). If offered, the worker chooses a level of training investment in customizing the AI system. A higher level of training investment may lower the worker's cost of effort, may improve the likelihood of better work outcomes, or both. After observing the worker's training investment into customizing the AI system, the manager offers the worker a contract, the worker decides on their effort level, and outcomes and payoffs are realized. 

\begin{table}
	\caption{Task verifiability and editability: a summary of some prior empirical findings}
	\label{tab:one}
	\begin{minipage}{\columnwidth}
		\begin{center}
			\begin{tabular}{l|l l}
				\toprule
                & Easy to verify & Hard to verify \rule{0pt}{2.6ex} \rule[-1.8ex]{0pt}{0pt} \\
                \hline
				Easy to edit & 
                \begin{minipage}[t]{0.4\columnwidth}
                \emph{Example}: Customer service, writing\\[1ex]
                \emph{Outcome}: Lower-skilled workers saw a greater quality boost compared to high-skilled workers in customer service.
                \end{minipage} & 
                \begin{minipage}[t]{0.4\columnwidth}
                \emph{Example}: Entrepreneurship advice, financial modeling\\[1ex]
                \emph{Outcome}: High-skilled workers saw quality improvement when provided AI advice for entrepreneurship, whereas low-skilled workers saw worse performance.
                \end{minipage} \rule{0pt}{3ex} \\
                Hard to edit & 
                \begin{minipage}[t]{0.4\columnwidth}
                \emph{Example}: Software maintenance, art\\[1ex]
                \emph{Outcome}: Fast adoption among unskilled workers (i.e. non-artists) at the risk of professional artists.
                \end{minipage} & 
                \begin{minipage}[t]{0.4\columnwidth}
                \emph{Example}: Software feature development, mathematical proofs\\[1ex]
                \emph{Outcome}: No increased improvement observed for unskilled open-source contributors.
                \end{minipage} \rule{0pt}{5ex} \rule[-15ex]{0pt}{0pt} \\
				\bottomrule 
			\end{tabular}
		\end{center}
		\bigskip\centering
		\footnotesize\emph{Source:} \cite{Yeverechyahu_2024_ImpactLargeLanguage,Brynjolfsson_2023_GenerativeAIWork,Otis_2024_UnevenImpactGenerative}
	\end{minipage}
\end{table}

Our results from a single period model show that a strategic worker will invest in training their AI twin only when this training increases the AI twin's efficacy growth \textit{faster} than that of the human and AI working in conjunction and will underinvest in other settings, trading off improvements in output quality and ease of effort to preserve their wage bargaining power and mitigate their risk of displacement. We explain how the growth pattern that incentivizes investment increases the information rent a worker commands by lowering \textit{outcome separability}---the extent to which the manager can map observed outcomes to worker effort---and how it is associated with high levels of verifiability. 

A strategic worker's incentives to improve their AI twin are constrained by two additional factors. Reducing the cost of effort, which is associated with a greater level of editability, has the additional effect of lowering a worker's wage bargaining power. Thus, higher levels of training investment are only incentive-aligned for work with relatively low editability. Additionally, a worker's training investments are sometimes constrained by the risk that the AI twin's performance becomes sufficiently good to cause the worker to be displaced entirely. We show this risk is greater when the task has low \textit{quality importance}, or when there is only weak benefit to the manager from good instead of bad outcomes.

We build on these results to examine, within a two-period model, how the behavior of non-strategic (or myopic) workers might deviate from those of a strategic worker. We characterize the conditions under which such agents might overinvest into training their AI twins and then get displaced by them, and show that these conditions are more likely to occur in settings where the training investments lead to greater immediate returns that allow the myopic worker to "shirk" in the short term. Thus, the settings in which training investments yield the highest immediate personal productivity gains also put the agent at the greatest risk of mistakenly creating an AI twin that will displace them in the long run. However, we also find that in some subset of these conditions, when the efficiency gains for the AI twin alone are high enough, even a strategic agent may opt for short-term effort reduction gains with a higher future risk of displacement. 

The results and framework presented in this paper offer a new lens through which  individual investments into improving generative AI systems might be viewed. A hallmark of our framework is that it does not make assumptions about the nature of the AI implementation or work it supports. As generative AI continues to evolve beyond current large language models and towards more sophisticated multimodal systems and advanced reasoning capabilities, we believe our simple framework will remain relevant. 

We show that the tradeoffs associated with a strategic human’s incentives to actively contribute to the creation of better generative AI systems are real. Even when one's AI twin enables the production or higher-quality output or a reduction in effort, incentives to improve this twin are often constrained by the risk of replacement or the potential lowering of earnings. While counterintuitive at first glance, the finding --- that training an AI twin is incentive-aligned only when the stand-alone gains grow faster than those when the human and AI working in conjunction --- makes sense when one considers a worker's desire to preserve their wage bargaining power. 

Further, our finding that the strategic imperative a human might have to contribute less to the AI twin highlights the importance of careful incentive alignment when aiming to induce human workers to contribute to the improvement and customization of generative AI systems. As our two-period model shows, in the short run, workers may mistakenly embrace and overinvest in these training efforts, but once it becomes clear that those investments that yield the highest immediate personal productivity gains also put one at the greatest risk of mistakenly creating an AI twin that will displace them in the long run, this behavior will change. 

Our ongoing work is focused on extending and generalizing our model while also refining its underlying framework. Our core results generalize naturally to a setting involving continuous outcomes where more training leads to an outcome distribution that is better in the sense of first-order stochastic dominance. However, this extension also allows us to define the nature of the work and the technology in a more nuanced manner by expanding the ways in which we can differentiate the distribution of outcomes of the stand-alone AI twin from the human-AI combination, considering complementarity versus substitution and also allowing a richer specification of verifiability that separates the stand-alone AI twin's outcome distribution from that of the human-AI combination even when the human exerts low effort.  

Reinterpreted appropriately, our results could also suggest that effective AI tools will incorporate features that make outputs more transparent and assessable by non-experts, while also providing robust editing capabilities that allow for meaningful human intervention. This interpretation provides another argument in favor of a shift away from black-box solutions toward more interpretable and malleable AI systems. Strengthening this kind of conclusion from our analytical results will require an approach that unpacks the connection between the tasks comprising the work being done and the capabilities of the supporting AI, and models them separately. 

The empirical literature connecting generative AI to productivity changes is growing rapidly. As we encounter new results that deepen our understanding of the nature of interaction between generative AI and human workers, it is possible that other aspects of work beyond editability and verifiability could be relevant. 

As the returns from improvements to general-purpose generative AI systems are increasingly constrained by the plateauing of "scaling laws," by the unavailability of increasingly larger training data sets, and by the specter of "model collapse," engaging human workers in the improvement of AI systems will be increasingly important for organizations to continue to realize productivity gains from generative AI. The realization that the recent performance improvements of DeepSeek over OpenAI's systems stem from not just the use of more sophisticated reasoning but from a heavier reliance on RLHF only underscore this importance. 

Preserving and sustaining improvements in company-specific generative AI will likely require nuanced design of human incentives. The viability of contracting that align a human's returns from contributing to improvements in AI through, for example, fractional IP ownership, may be countered by the associated transaction costs and challenges in attribution, but this nevertheless represents a  direction for future inquiry. Policy interventions that better align an organization's incentives to balance worker returns with generative AI improvements could  yield more reliable training inputs and more sustained long-run productivity gains for a country. 

\section{A single period model}
We use a two-outcome, limited-liability, hidden action model. A manager (principal) wishes to hire a worker (agent) to perform a task. Both parties are risk neutral. The agent chooses between two effort levels \(e\), high (\(e=1\)) and low (\(e=0\)). A later section will analyze the generalization to a continuous, multi-effort setting. Exerting high effort involves the agent incurring a cost \(c>0\). We discuss the relationship between cost of effort and the use of AI later in this section. The principal's payoff is shaped by outcomes that are realized from the agent's exertion of effort. 

We follow the standard approach to modeling this setting of moral hazard. If the principal could observe the agent's effort, either directly or by inferring it perfectly from observable output, they would pay the agent conditional on the effort observed (the "first-best" setting). When the principal cannot observe the agent's effort, the observed work outcome depends probabilistically on the agent's effort and the principal can only pay the agent based on outcome rather than on effort (the "second-best" setting). 

We assume two possible outcomes that, respectively, provide benefit \(\bar{S}\) and \(\underline{S}\) to the principal, where \(\bar{S}>\underline{S}\). We designate \(\pi_1\) as the probability that high effort produces the better outcome, and \(\pi_0\) as the probability that low effort produces the better outcome. We reinterpret these probabilities in the context of AI usage later in this section. The agent's outside option is set to 0.\footnote{In the language of contract theory, the agent has "limited liability," meaning they cannot pay more than \(L\) to the principal. To simplify the analysis, we assume that \(L=0\). } 

Since the principal can observe the agent's outcome but not their effort, the contract offered by the principal will depend on the outcome observed. Denote by \(\bar{t} ,\, \underline{t}\) the payments to the agent for the better (\(\bar{S}\)) and worse (\(\underline{S}\)) outcomes respectively. We analyze contracts where \(\bar{t} > \underline{t}\), since if \(\bar{t} \le \underline{t}\), the agent will choose effort \(e=0\). 

Additionally, the principal can choose to offer the agent a customizable generative AI system that we refer to as a \textit{trainable AI twin}. The level of capability of the trainable AI twin can be improved by the agent through an investment \(v \geq 0\), which in this setting is indicative of the amount of training data and fine-tuning feedback the agent provides. The lack of investment \(v=0\) reflects the agent's baseline capabilities, or the absence of an AI twin. 

To focus our analysis on the strategic aspects of the agent's choices, we assume that this training investment \(v\) involves no direct cost to the agent or the principal. Examples of this kind of investment could involve the agent providing examples of their writing style to fine-tune an LLM, the agent annotating data about their decisions and communications to improve an AI system's understanding of them, or the agent participating in rating the quality of a system's output in response to their prompts for use in reinforcement learning with human feedback (RLHF). 

The choice of \(v\) may alter the agent's cost of effort, now denoted \(c(v)>0\), as well as the relative likelihoods of better and worse outcomes, now denoted \(\pi_0(v)\) and \(\pi_1(v)\) respectively. These values are all continuous in \(v\). A larger training investment \(v\) weakly increases the likelihood of better outcomes, and weakly lowers the agent's cost of exerting effort once the AI twin is trained, and thus, \(\pi_1'(v) \geq 0,\, \pi_0'(v) \geq 0,\, c'(v) \leq 0\). We return to interpretations of these rates of change, whose relative values are central to our results, later in this paper. 

With the addition of an AI twin, one can now think of \(e=0\) as characterizing the ``AI only" setting, where \(\pi_0(v)\) is the outcome distribution induced by AI without additional human effort, and \(e=1\) as characterizing the ``human+AI" setting, where \(\pi_1(v)\) is the outcome distribution induced when human effort complements the AI twin. Importantly, we are considering a setting in which a higher level of training investment in the AI twin can also improve its \textit{standalone} performance. 

It is natural to assume that the standalone AI performance is always weakly inferior to the setting where the human exerts high effort and collaborates with the AI, or that \(\pi_1(v) > \pi_0(v)\) for all \(v\). Let there be some \(v_{\max } >0\) that bounds how much an agent can invest in their AI twin. This limit could be the result of a predetermined model size --- the performance and sample efficiency of a language model scales with the size of the model \cite{Kaplan_2020_ScalingLawsNeural} --- or because the size of the training data corpus is limited by the agent's productive capacity. 

Before proceeding further, we note that since \(\pi_1'(v)>0\), \(\pi_0'(v)>0\) and \(c'(v)<0\), the socially optimal level of training investment is \(v=v_{\max}\). 

\begin{lemma}
    \label{lma:soc-opt}
    The socially optimal level of training investment is \(v=v_{\max}\). 
\end{lemma}

\subsection{Optimal contract}
The principal decides to provide the AI twin if, taking into account the anticipated training investment $v$ by the agent and the ensuing choices of effort, the presence of the AI twin increases the expected surplus of the principal. 

Assuming they provide an AI twin, for a fixed level of training investment \(v\), the principal will choose a menu of payments \(\{\bar{t}(v), \underline{t}(v)\}\) to maximize their surplus:
\begin{equation}
   \label{eqn:mgr-prob}
  \max_{\{\bar{t}(v), \underline{t}(v)\} } \pi_1(v) (\bar{S}-\bar{t}(v)) + (1-\pi_1(v)) (\underline{S} - \underline{t}(v)).
\end{equation}
In order to induce both participation and high effort from the agent, the principal's payments must satisfy the agent's participation and incentive compatibility constraints, respectively: 
\begin{eqnarray}
    \label{eqn:wrk-ir}
    \pi_1(v) \bar{t}(v) + (1-\pi_1(v)) \underline{t}(v) - c(v) \geq 0,\\
    \label{eqn:wrk-ic}
    \pi_1 \bar{t}(v) + (1-\pi_1(v)) \underline{t}(v) - c \geq \pi_0(v) \bar{t}(v) + (1-\pi_1(v)) \underline{t}(v).
\end{eqnarray}

The optimal contract follows from standard textbook contract theory \citep{Laffont_2002_TheoryIncentivesPrincipalAgent} dealing with hidden action settings that have two effort levels and two outcomes: 
\begin{lemma}
    \label{lma:optimal-contract}
    Given a fixed level of investment \(v\) by the agent, the optimal contract offered by the principal is \(\bar{t}(v) = \frac{c(v)}{\pi_1(v) - \pi_0(v)},\, \underline{t}(v) = 0\). 
\end{lemma}

(A proof of this lemma is available on request.) 

Under this contract, the principal has an expected surplus of 
\begin{equation}
    \label{eqn:mgr-surplus0}
    \pi_1 (\bar{S}-\frac{c(v)}{\pi_1(v) - \pi_0(v)}) + (1-\pi_1(v)) \underline{S}, 
\end{equation}
or of
\begin{equation}
    \label{eqn:mgr-surplus}
    \pi_1(v) \bar{S} + (1-\pi_1(v)) \underline{S} - \frac{\pi_1(v)}{\pi_1(v) - \pi_0(v)}c(v). 
\end{equation}

The principal thus offers a contract to induce a high level of effort if the incremental expected payoff from a higher effort level versus a lower effort level is at least as high as the expected payment to the agent, or if
\begin{equation}
\label{eqn:induce-effort}
    (\pi_1(v) - \pi_0(v)) (\bar{S} -\underline{S}) \geq \pi_1(v) \frac{c(v)}{\pi_1(v) - \pi_0(v)}.
\end{equation}

The agent's expected surplus is 
\begin{equation}
    \label{eqn:wrk-surplus}
    U(v)=\pi_1(v)\frac{c(v)}{\pi_1(v) - \pi_0(v)} + (1-\pi_1(v))0 - c(v) = \frac{\pi_0(v)}{\pi_1(v) - \pi_0(v)}c(v).
\end{equation} 

We focus on settings for which, in the absence of an AI twin, the principal wishes to offer a contract to induce high agent effort, or that 

\begin{equation}
\label{eqn:induce-effort-no-investment}
    (\pi_1(0) - \pi_0(0)) (\bar{S} -\underline{S}) \geq \pi_1(0) \frac{c(0)}{\pi_1(0) - \pi_0(0)}.
\end{equation}

Put differently, we imagine this baseline as the pre-AI status quo, and assume that in this status quo, there are gains from contracting between the principal and the agent that can be implemented. 

\subsection{AI twin: Incentives for both parties}
\begin{figure}
    \centering
    \includegraphics[width=0.6\linewidth]{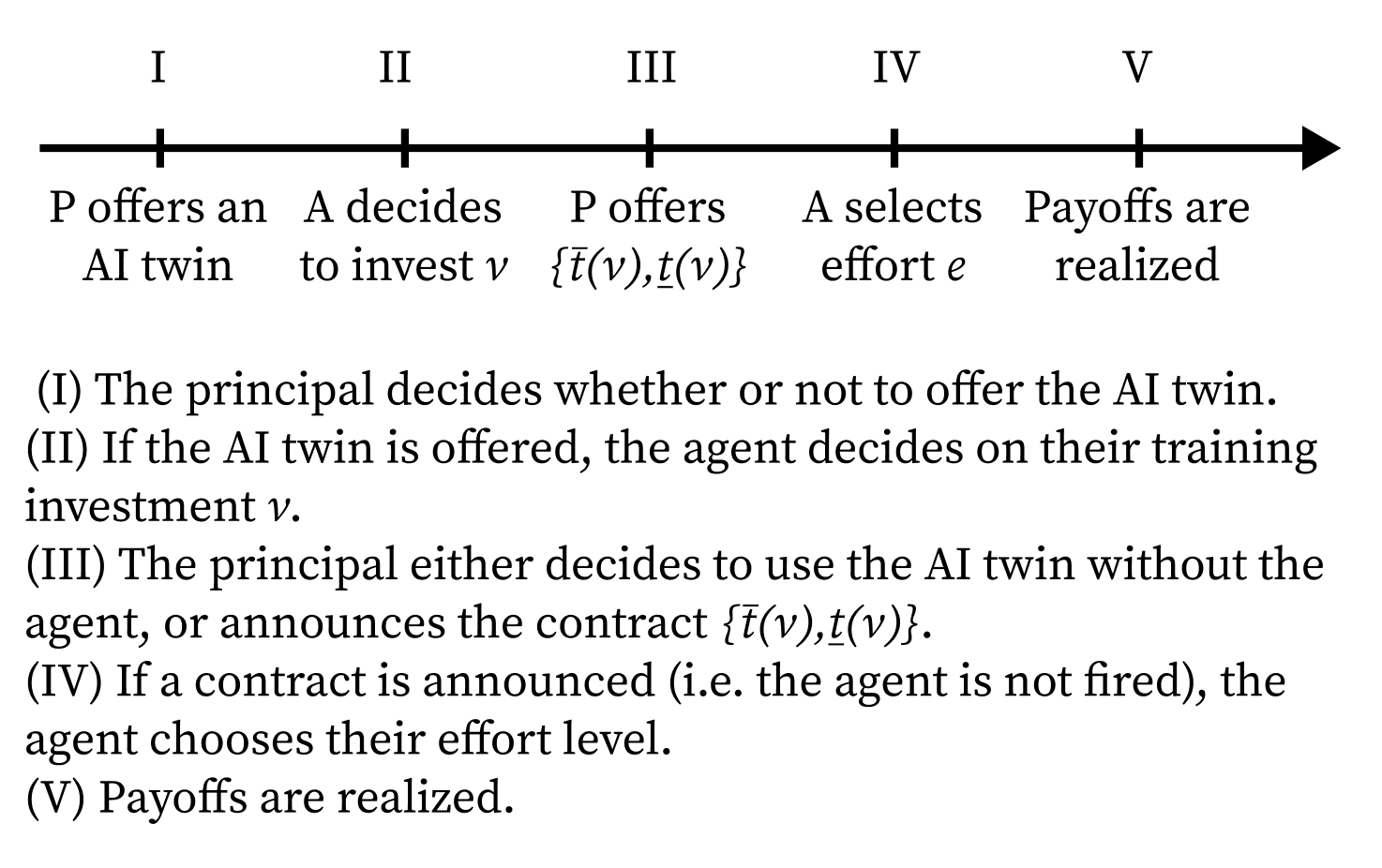}
    \caption{Timing in the basic model}
    \label{fig:single-timeline}
\end{figure}

We now consider the agent's incentives for making a training investment into their AI twin and the principal's incentives for offering this AI twin. The sequence of events is as follows (See Figure \ref{fig:single-timeline}): 

(I) The principal decides whether or not to offer the AI twin. 

(II) If the AI twin is offered, the agent decides on their training investment \(v\). 

(III) The principal either decides to use the AI twin without the agent or announces the contract \(\{\bar{t}(v),\, \underline{t}(v)\}\). 

(IV) If a contract is announced (that is, the agent is not fired), the agent chooses their effort level. 

(V) Payoffs are realized. 

Given this sequence of events, assuming step (IV) is reached, the contract structure will follow Lemma \ref{lma:optimal-contract}:
\begin{equation} 
    \bar{t}(v) = \frac{c(v)}{\pi_1(v) - \pi_0(v)}, 
    \underline{t}(v) = 0,
\end{equation}
and following (\ref{eqn:wrk-surplus}), the agent's expected surplus is \(U(v)\).

Now, assuming step (III) is reached, the principal will offer a contract only if, anticipating inducing a high effort level from the agent in step (IV), their expected surplus from doing so is higher than their expected surplus from using the AI twin imbued with the agent's training investment \(v\) but without the agent involved in exerting effort, or if: 
\begin{equation} 
    \label{eqn:mgr-ic}
    \pi_1(v) (\bar{S}-\bar{t}(v)) + (1-\pi_1(v)) (\underline{S} - \underline{t}(v)) \geq \pi_0(v) \bar{S} + (1-\pi_1(v)) \underline{S}.
\end{equation}

Note that (\ref{eqn:mgr-ic}) above also specifies a constraint on the agent's level of training investment. A strategic agent will not choose a training investment level that reverses the inequality in (\ref{eqn:mgr-ic}). This is because a strategic agent does not want to be displaced by their AI twin and lose the positive expected surplus associated with exerting high effort. For this reason, in what follows, we sometimes refer to \eqref{eqn:mgr-ic} as the "displacement deterrent" condition.

Anticipating the principal's choices in step (III) to induce effort, the agent will decide an investment level \(v\) in step (II) to maximize their surplus \(U(v)\). Finally, in step (I), the principal offers the AI twin if doing so increases the principal's surplus, anticipating offering a contract and inducing a high level of effort from the agent:
\begin{equation}
    \label{eqn:mgr-ir}
    \pi_1(v) (\bar{S}-\bar{t}(v)) + (1-\pi_1(v)) (\underline{S} - \underline{t}(v)) \geq \pi_1(0) (\bar{S}-\bar{t}(0)) + (1-\pi_1(0)) (\underline{S} - \underline{t}(0)).
\end{equation}
\subsection{AI twin: Agent's training investment}
With these preliminaries out of the way, we are now ready to analyze the agent's choice of training investment in step (II). The agent chooses a training investment level that solves:
\begin{equation}
    \label{eqn:wrk-invest}
    \max_{v \in [0,v_{\max}]} U(v) = \frac{\pi_0(v)}{\pi_1(v) - \pi_0(v)}c(v)
\end{equation}
subject to \eqref{eqn:mgr-ic}.  The objective function \(U(v)\) in (\ref{eqn:wrk-invest}) is not necessarily strictly quasiconcave in \(v\). Rather than imposing additional structure through assumptions that would ensure an interior solution, we instead focus on deriving and interpreting conditions under which \eqref{eqn:wrk-invest} has a well-defined solution, which may either be an interior solution or one of the end points.

\subsubsection{No displacement deterrent}
We begin by ignoring the displacement deterrent condition, which we consider in the next subsection, and focus on the conditions that lead the agent to choose not to invest in their AI twin, to choose the maximum training investment, and to choose an intermediate level of investment. 
Define \( Q_\pi(v) \), the \textit{outcome separability} of the work in question, as:
   \begin{equation}
    \label{eqn:separable}
    Q_\pi(v) = \frac{\pi_1(v) }{\pi_0(v)}.
   \end{equation}
We call \( Q_\pi(v) \) outcome separability because it measures the confidence that the principal has, after observing the realized outcome, in their assessment of the agent's effort level. Also note that 

   \begin{equation}
    \label{eqn:equiv}
    Q_\pi(v) -1 = \frac{\pi_1(v) - \pi_0(v)}{\pi_0(v)},
   \end{equation}
and thus, 
   \begin{equation}
    \label{eqn:new-u}
    U(v) = \frac{c(v)}{Q_\pi(v) -1}.
   \end{equation}
As (\ref{eqn:new-u}) suggests, and the following results show, the optimal level of training investment \(v\) made by the agent will depend on the relative rate at which the training investment alters outcome separability and the agent's cost of effort. 

\begin{proposition}
    \label{prop:no-inv}
    If, for all \(v \in [0,v_{\max}]\),
    \begin{equation}
        \frac{c'(v)}{c(v)} < \frac{Q_\pi'(v)}{Q_\pi(v)-1}
    \end{equation} 
    then the agent's optimal training investment is \(v=0\).
\end{proposition}

Proofs are available in the Appendix. 

Since \(\pi_1(v) - \pi_0(v)>0\), it follows that \(Q_\pi(v)-1 > 0\), and since \(c'(v)\leq 0\), a natural corollary is that when \(Q_\pi'(v)>0\) for all \(v\), the agent will choose \(v=0\). 

Next, we characterize when the agent chooses the maximum training investment. 

\begin{proposition}
    \label{prop:max-inv}
    If, for all \(v \in [0,v_{\max}]\), \(Q_\pi'(v) \leq 0\) and 
    \begin{equation}
        \bigg \lvert \frac{c'(v)}{c(v)} \bigg \rvert < \bigg \lvert \frac{Q_\pi'(v)}{Q_\pi(v)-1} \bigg \rvert
    \end{equation} 
    then the agent's optimal training investment is \(v=v_{\max}\).
\end{proposition}

Before characterizing a set of sufficient conditions for an optimal interior training investment choice \(v \in [0,v_{\max}]\), it is helpful to interpret Proposition \ref{prop:no-inv} and Proposition \ref{prop:max-inv} based on the framework we introduced in Section 2. 

In our setting, the agent is the functional ``expert" --- when presented with output from the AI twin, they are able to assess its quality and, if needed, make changes. One can think of the rate at which the cost of effort goes down with training investment, or the absolute value \(c'(v)\), as characterizing the \textit{editability} of the task: a greater reduction in the cost of effort implies that the AI output can easily be modified to increase the likelihood of a good outcome. 

In addition, the rate at which outcome separability changes, the value of \(Q_\pi'(v)\), characterizes the \textit{verifiability} of the work in question. Suppose for a moment that as the training investment increases, ``human+AI" outcomes improve faster than standalone AI outcomes, or \(\pi_1'(v) > \pi_0'(v)\), which in turn implies that \(Q_\pi'(v)>0\). This corresponds to a situation in which, absent the human worker's expertise, the principal has a lower ability to discern if the AI twin's output will lead to high quality. By contrast, if the principal is able to appropriately interpret and deploy the AI twin's output without the help of the agent, the standalone AI outcomes improve faster than ``human+AI" outcomes. The task's verifiability is therefore modeled by the rate of change in outcome separability: Outcome separability increasing with training investment corresponds to low verifiability and a sufficiently negative value of \(Q_\pi'(v)\) corresponds to high verifiability. 

The results of Proposition \ref{prop:max-inv} show that the relative rate at which outcome separability decreases has to be sufficiently larger than the corresponding gains from a lower cost of effort for the agent to be willing to make a training investment. Thus, the incentives for the agent to make training investments are greatest for work in which both editability is sufficiently low and verifiability is sufficiently high. For example, an investment banking analyst, if given an AI system that can automate steps in their valuation or other financial modeling, is more likely to invest in improving this system if they are still needed to assess and vouch for whether the model is accurate.

\begin{proposition}
    \label{prop:interior}
    A set of sufficient conditions for the agent to choose an interior value of \(v \in (0,v_{\max})\) are:
    \begin{equation}
        \frac{c'(0)}{c(0)} > \frac{Q_\pi'(0)}{Q_\pi(0)-1},
     \end{equation} 
     and
     \begin{equation}
        \frac{c'(v_{\max})}{c(v_{\max})} < \frac{Q_\pi'(v_{\max})}{Q_\pi(v_{\max})-1}.
    \end{equation} 
\end{proposition}

A more restrictive sufficient condition is that for all \(v \in [0,v_{\max}]\), \(Q_{\pi}'(v) < 0\), \(Q_{\pi}''(v) > 0\) and \(Q_{\pi}''(v) > c''(v)\), which would imply the conditions of Proposition \ref{prop:interior}. An illustration of the different regions of the parameter space for which the conditions of our three propositions hold is provided in Figure 2. Note that since the agent always invests \(v=0\) when \(Q_{\pi}'(v)>0\), outcomes are illustrated for \(Q_{\pi}'(v) \leq 0\). 

\begin{figure}
    \centering
    \includegraphics[width=0.8\linewidth]{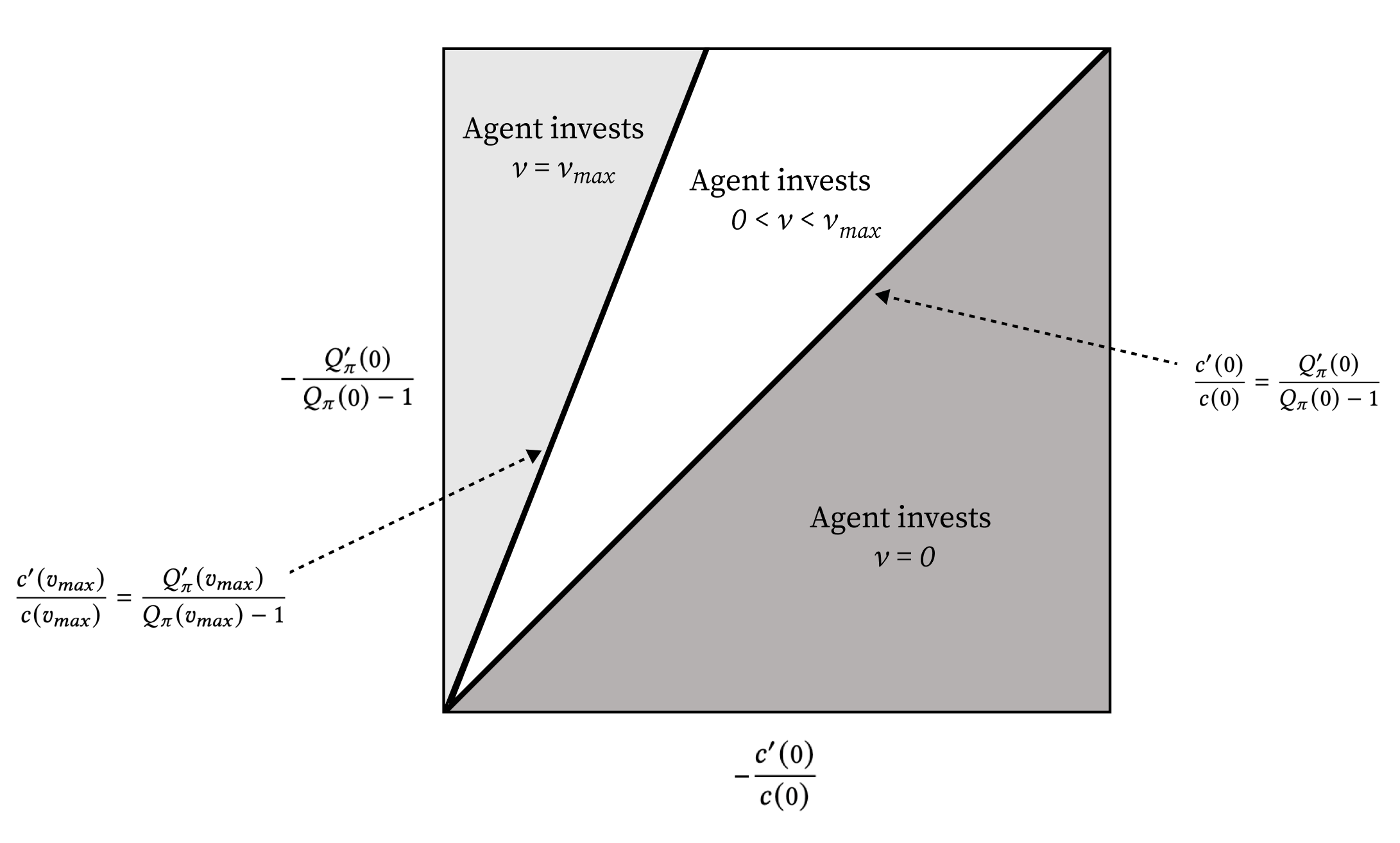}
    \caption{An illustration of optimal training investment regions as $c(v)$ and $Q_\pi(v)$ vary.}
    \label{fig:fig2}
\end{figure}

One might intuitively think that if training investments lower an agent's cost of effort sufficiently, the agent would embrace the prospect of an AI twin and actively make these investments. However, through this lens, it is clear that the agent's incentives to make training investments are instead shaped by trading off a higher likelihood of a large payoff (which incentivizes a higher \(v\)) against preserving higher wages by maintaining a sufficiently high cost of effort and by lowering outcome separability. These investments lower the agent's ability to command a higher wage, and as we show in Section 4, a non-strategic agent who makes this kind of investment finds their earning potential harmed in future periods even if they are not entirely displaced by the AI twin. Paradoxically, therefore, when the AI twin can ease the burden on the human worker more, the worker has less incentive to improve the AI twin, and this further bounds the competing trade-off of maintaining ``unobservability" of effort by keeping outcome separability relatively low. 

\subsubsection{Displacement as a deterrent to making a training investment}
We now return to the possibility that the displacement deterrent condition (\ref{eqn:mgr-ic}) leads the agent to invest below the levels specified by Propositions \ref{prop:max-inv} and \ref{prop:interior}. 

Although \(\bar{t}(v)\) is not a primitive of our model but a derived function, its properties (and, in particular, whether it is increasing or decreasing) are a convenient way to characterize the results about when an agent is threatened by displacement by their AI twin. In what follows, the issue of displacement depends directly on the difference in the rates at which \(\pi_1(v)\) and  \(\pi_0(v)\) increase (rather than on their ratio \(Q_{\pi}(v)\)), and so we define
\begin{equation}
    \Delta\pi(v)=\pi_1(v)-\pi_1(v).
\end{equation}
From Lemma \ref{lma:optimal-contract}, it follows that 
\begin{equation}
    \label{eqn:t}
    \bar{t}(v) = \frac{c(v)}{\Delta\pi(v)}.
\end{equation}

The next lemma follows immediately from \eqref{eqn:t} and the fact that \(c'(v)<0\):
\begin{lemma}
    \label{lma:t}
    The following conditions partially characterize when \(\bar{t}(v)\) is increasing or decreasing: 
    \begin{itemize}
       \item If \(\Delta\pi'(v)>0\), then \(\bar{t}'(v)<0\). 
       \item \(\bar{t}'(v)>0\) only if \(\Delta\pi(v)\) is decreasing and decreases sufficiently faster than \(c(v)\). 
    \end{itemize}
\end{lemma}

Furthermore, $\bar{t}'(v) > 0$ is a sufficient condition for Proposition \ref{prop:max-inv}: 

\begin{lemma}
    \label{lma:increasing-t}
    If $\bar{t}'(v) > 0$, then $Q_\pi'(v) < 0$ and $\bigg \lvert \frac{c'(v)}{c(v)} \bigg \rvert < \bigg \lvert \frac{Q_\pi'(v)}{Q_\pi(v)-1} \bigg \rvert$.
\end{lemma}

Now, using (9) and (13), the displacement deterrent condition can be rearranged as follows:
\begin{equation}
    \label{eq:dd}
    (\bar{S}-\underline{S})\left(1-\frac{1}{Q_\pi(v)}\right) \geq \bar{t}(v).
\end{equation}
We refer to \((\bar{S} - \underline{S})\) as the \textit{quality importance} of the work being done, since it measures the difference in payoff to the principal from  high-quality and low-quality output. The quality importance represents how much is at stake for the principal in accomplishing the task well.

Since \(0 < 1 - \frac{1}{Q_{\pi}(v)} < 1\), the left-hand side of \eqref{eq:dd} can never be greater than \((\bar{S}-\underline{S})\). Thus, if \(\bar{t}'(v) \geq 0\) and \(\bar{t}(v_{\max}) > (\bar{S}-\underline{S})\), because \(\bar{t}(v)\) is continuous, then there will be some threshold value \(v=v^*\) implicitly defined by

\begin{equation}
    \label{eq:v*}
    (\bar{S}-\underline{S})\left(1-\frac{1}{Q_\pi(v^*)}\right) = \bar{t}(v^*)
\end{equation}

such that the human will be displaced by their AI twin for any training investment \(v > v^*\). We refer to \(v^*\) as the \textit{displacement threshold}. 

When is it likely that this displacement threshold will constrain an agent's training investment? Recall from the corollary to Proposition \ref{prop:no-inv} that if \(Q_\pi'(v)>0\) , the agent makes no training investment. It is easy to verify that \(Q_\pi'(v)>0\) implies \(\Delta\pi'(v)>0\), and \(\Delta\pi'(v)<0\) implies that \(Q_\pi'(v)<0\). Thus, the more relevant case for us to examine is the second setting of Lemma \ref{lma:t}, when \(\bar{t}'(v)>0\), since the setting where \(\bar{t}'(v)<0\) is more likely to be associated with the agent making no training investment anyway, which would occur, for example, when the work is verifiable but not editable. 

When \(\bar{t}'(v)>0\), the left-hand side of \eqref{eq:dd} is more likely to bind the inequality when the quality importance \((\bar{S} - \underline{S})\) of the work is lower. For example, quality matters much less for a meme on a blog post than for a painting being auctioned at Christie's. Quality importance also grows with how harmful the bad outcome is---for example, relying entirely on a robotic surgeon that underperforms a human surgeon even slightly may only lower the likelihood of success a little, but the consequences of a bad outcome are substantial. Similarly, a digital programming twin that is empowered to autonomously edit the codebase of an entire application could introduce bugs that, if not detected, have severe damaging effects. 

Put differently, the risk of displacement is inversely proportional to how important it is to perform the task well. When the stakes are lower, our measure of quality importance \((\bar{S} - \underline{S})\) is lower, and hitting the displacement threshold becomes more likely. Current cases of generative AI-enabled automation seem to support this. Art, particularly as used in background designs, is one class of jobs in which replacement is occurring at a rapid pace \citep{Demirci_2025_WhoAIReplacing}. Using current image generation tools, a marketing manager can quickly create an emotionally appealing image, or a blogger can use an LLM create a vaguely inspirational quote to break up walls of text. Such low-stakes tasks do not need the level of precision that a skilled worker can contribute, and thus are commonly delegated to AI. Many other types of work that are listed on freelance websites may be considered low-stakes by the lister, making the workers on these marketplaces especially susceptible to AI displacement. Conversely, this inequality also predicts that humans engaged in high-stakes tasks face a lower risk of displacement and, assuming it is otherwise incentive-aligned, can invest more readily in their AI twins. 

Our analysis thus far assumes that agents are sufficiently strategic to anticipate the consequences of their training investments. However, it is possible that today's workers are making training investments to reduce their cost of effort or improve outcomes without fully considering the future consequences of their present-day choices. We explore some aspects of this scenario in a two-period extension of our model.

\section{Two-period model with a myopic agent}

In this section, we consider a two-period extension, contrasting the behavior of a myopic agent who does not internalize the future consequences of their training investment choices with that of the single-period strategic agent. We also explore extensions which assess the effects of the time-persistence the learning of the AI twin, and the generalization of the model to continuous effort settings.

\begin{figure}
    \centering
    \includegraphics[width=0.9\linewidth]{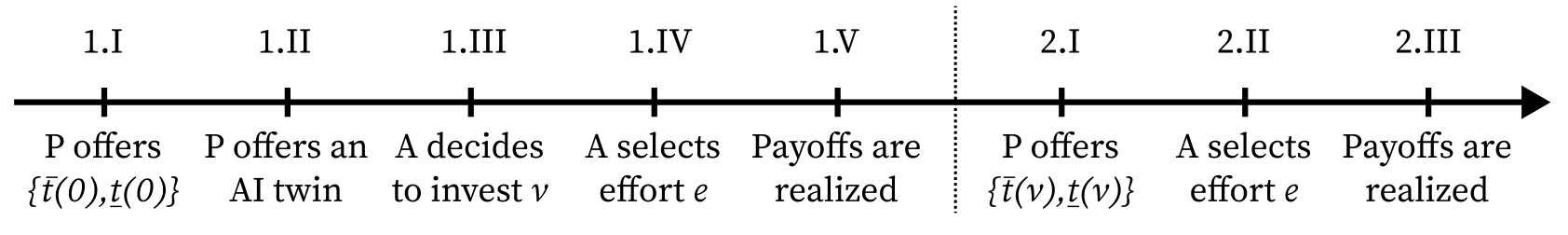}
    \caption{Timing of two-period model}
    \label{fig:two-step}
\end{figure}

The specific setting we model is summarized in the timeline of Figure \ref{fig:two-step}. In the first period, the principal offers a contract. After this contract is offered, the principal can offer the AI twin to the agent, but cannot change the contract.  The agent then chooses their training investment level \(v\). After that, the agent chooses their effort level, and first-period payoffs are realized. 

In the second period, the principal, taking into account the training investment made by the agent, decides whether to offer a new contract to the agent, or to simply use the AI twin instead. If the contract is offered, the agent cannot alter their training investment (the AI twin has already been trained), and only decides whether to accept or reject the contract. If the contract is accepted, the agent chooses an effort level, and payoffs are realized. We begin with a baseline two-period model, and then briefly examine the implications of the AI "degrading" over time, and of the second period by itself comprising an infinite horizon of periods. 

Denote the training investment made by the myopic agent as \(v^M\). Since the agent is myopic, they only consider their first period payoff when deciding their investment level. Recall as well that the agent now chooses their training investment after the principal proposes a contract but before they decide their effort. Thus, it is possible for the training investment following the contract offer to create a situation where it is optimal for the agent to "shirk," or to choose the lower rather than higher effort level. 

We first establish that independent of their subsequent period 1 choice of effort, the agent will always choose the maximum training investment. Consider any contract $\{\bar{t},\underline{t}\}$ offered by the principal in period 1. If the agent intends to exert the higher effort level, then:
\begin{equation}
 \label{eqn:M1}
    v^M = \arg\max_{v \in [0,v_{\max}]} \pi_1(v) \bar{t} + (1-\pi_1(v))\underline{t} - c(v). 
\end{equation}
Since \(\pi_1'(v) > 0\) and \(c'(v)<0\), the objective function in  \eqref{eqn:M1} is increasing in \(v\) and thus, \(v^M=v_{\max}\). Similarly, if the agent intends to exert the lower effort level, then 
\begin{equation}
    \label{eqn:M2}
    v^M = \arg\max_{v \in [0,v_{\max}]} \pi_0(v) \bar{t} + (1-\pi_0(v))\underline{t}, 
\end{equation}
and since \(\pi_0'(v) > 0\), the objective function in \eqref{eqn:M2} is increasing in \(v\) and thus, \(v^M=v_{\max}\). Since the principal's offer of a contract is fixed before the agent chooses to invest, the myopic agent's training investment is independent of the principal's offer. 

When will the agent's training investment in their AI twin cause them to "shirk"? Comparing \eqref{eqn:M1} and \eqref{eqn:M2}, this occurs when 
\begin{equation}
    \label{eqn:shirk}
    \pi_1(v_{\max}) (\bar{t}-\underline{t}) - c(v_{\max}) < \pi_0(v_{\max}) (\bar{t}-\underline{t}).
\end{equation}
Rearranging \eqref{eqn:shirk}, this condition reduces to:
\begin{equation}
    \label{eqn:shirk2}
    \bar{t}-\underline{t}<\frac{c(v_{\max})}{\pi_1(v_{\max})-\pi_0(v_{\max})}.
\end{equation}
Following Lemma \ref{lma:optimal-contract}, it follows that $\underline{t} = 0$. Thus \eqref{eqn:shirk2} implies that we have proved the following:
\begin{lemma}
    \label{lma:shirk}
    The myopic agent will "shirk" in period 1 after making the maximum training investment if the principal's offer is less than the optimal contract for the current level of investment, or if
    \begin{equation}
    \bar{t}<\bar{t}(v_{\max}). 
    \end{equation}
\end{lemma}

Next, we characterize when the myopic agent will get displaced by their AI twin. This follows directly from the definition of the displacement threshold defined in \eqref{eq:v*}. 

\begin{proposition}
    \label{prop:displace}
    If the displacement threshold \(v^* <v_{\max}\), then the myopic agent gets displaced by their AI twin in period 2. 
\end{proposition}

With this shirking condition in place, the principal must now decide between incentivizing the agent to not shirk by offering a contract at least as good as $\bar{t}(v_{\max})$, or choosing to induce the lower effort level.
The principal chooses the former if:
\begin{equation}
    \pi_1(v_{\max})(\bar{S} - \bar{t}(v_{\max})) + (1-\pi_1(v_{\max}))(\underline{S}-\underline{t}(v_{\max})) > \pi_0(v_{\max})\bar{S} - (1-\pi_0(v_{\max}))\underline{S},
\end{equation}

which is exactly the displacement deterrent condition \eqref{eq:dd}. Thus, it follows that both the first and second period contracts offered by the principal are determined by the contract derived in Lemma \ref{lma:optimal-contract}, under certain conditions: 
\begin{lemma}
    \label{lma:two-period}
    In the two-period model outlined above
    \begin{itemize}
        \item If the displacement threshold \(v^* < v_{\max}\), then the optimal contract for the principal in both periods is $\{0, 0\}$.
        \item If the displacement threshold does not exist or \(v^* \geq v_{\max}\), the optimal contract for the principal in both periods is \(\{\bar{t}(v_{\max}), \underline{t}(v_{\max})\}\). 
    \end{itemize}
\end{lemma}

The intuition for this result is simple: if the principal anticipates replacing the myopic agent after the first period, the same condition that makes this displacement optimal for the principal also makes it optimal to not induce a higher effort level in period 1. 

Notice that the condition that specifies when the agent will "shirk" in period 1 in Lemma \ref{lma:two-period} is satisfied if \(\bar{t}'(v)>0\), and recall from Section 3 that \(\bar{t}'(v)>0\) was also associated with a higher likelihood of a displacement threshold \(v^* < v_{\max}\). This connection seems natural --- if an agent decides that it is easier to use the AI twin and put in low effort in period 1, it stands to reason that a principal would reach the same conclusion in period 2.  When tasks are more verifiable, less editable and have a lower quality importance, this risk is especially significant. Ironically, therefore, the settings in which training investments yield the highest immediate personal productivity gains also put the agent at the greatest risk of mistakenly creating an AI twin that will displace them.

\section{Extensions}
\subsection{AI time persistence and infinite horizon}

We briefly discuss a couple of simple extensions. The first one considers the possibility that the AI twin's performance will degrade over time without additional training inputs, for example, because of new knowledge, shifting market conditions, changes in customer needs or the need for the AI to know about new products. Define $0 < \alpha < 1$ to be the time persistence of the AI twin. If the principal chooses to displace the agent with the AI twin in the second period, then absent the continued human input, the performance of the AI twin degrades, and the AI twin performs at the level associated with a lower training investment \(\alpha v\).
Consequently, the displacement deterrent condition in the second period changes accordingly:
\begin{equation}
    \label{eqn:displacement-degradation}
    \pi_1(v_{ \max})(\bar{S} -\bar{t}(v_{\max}))- (1-\pi_1( v_{\max}))(\underline{S}-\underline{t}(v_{\max})) > \pi_0(\alpha v_{ \max})\bar{S} - (1-\pi_0(\alpha v_{\max}))\underline{S}.
\end{equation}

The new displacement deterrent condition implies a new displacement level $v^\dagger(\alpha) > v^*$ under degradation. The principal will choose to employ the agent in period two if $v^\dagger(\alpha) > v_{\max}$. Although the agent may have chosen to shirk in period 1, in the scenario with AI twin degradation, if the degradation is large enough, condition \eqref{eqn:displacement-degradation} may be satisfied, and the principal may continue to employ the myopic agent in the second period. A lower time persistence thus leads to fragility in the AI twin that mitigates the risk faced by myopic agents who make poor training investment choices. This lends a note of caution to organizations that may opt hastily to fire their human workers only to regret it later when they face a degradation in the quality of their model outputs. 

Finally, we consider a setting where the second period is not a single period, but is a sequence of periods with future payoffs discounted at $\delta < 1$. 
In each of these periods, the principal can choose to offer a contract or not. As discussed before, if $v^* > v_{\max}$, the principal will always hire the agent. Assuming this condition is not met, it is easy to see that if this second period consists of sequence of $n$ periods, the displacement deterrence condition becomes:
\begin{equation}
\label{eq:infinite-degradation}
    \pi_1(v_{ \max})(\bar{S} -\bar{t}(v_{\max}))- (1-\pi_1( v_{\max}))(\underline{S}-\underline{t}(v_{\max})) < \pi_0(\alpha^n v_{ \max})\bar{S} - (1-\pi_0(\alpha^n v_{\max}))\underline{S}.
\end{equation}
As $n \rightarrow \infty$, $\alpha^n v_{\max} \rightarrow 0$. Thus, as the number of periods grows, at some point the performance of the AI twin would have degraded sufficiently to warrant rehiring the agent. Thus, one would expect a cyclical pattern where the principal hires the agent for one period to re-train the model, and then uses the AI twin for $m$ periods, and the cycle repeats.

\subsection{Continuous effort} 
For this next part, we use a continuous-effort, two-outcome model with limited liability. The agent can choose an effort \(e\) between \([\underbar{e}, \bar{e}] \), and produces an successful outcome with probability \(p(e)\). We say that \(p(e)\) is continuous, increasing and concave in \(e\), so that there is a single optimization. The cost of effort is given by \(c(e)\), which in this model we treat as linear in \(e\): \(c(e) = c_0 e\). The principal again receives \(\bar{S}\) in the successful outcome and \(\underbar{S}\) in the unsuccessful outcome. They will pay the agent \(\bar{t}\) in the case of a good outcome and \(\underbar{t}\) otherwise. The agent's utility is given by:

\[
U(e, \{\bar{t}, \underbar{t}\} ) = p(e) \bar{t} + (1 - p(e)) \underbar{t} - c(e) \]
The principal's utility is given by:
\[
U_P(e) = p(e) (\bar{S} - \bar{t}) + (1 - p(e)) (\underbar{S} - \underbar{t}) \]
The principal's problem is to choose the contract \((\bar{t}, \underbar{t})\) to maximize their utility, subject to the agent's individual rationality constraint and incentive compatibility constraint. The agent's individual rationality constraint is given by:
\[
U \geq U_0 \]
where \(U_0\) is the agent's reservation utility. We take \(U_0 = 0\). The incentive compatibility constraint is given by:
\[
e = \arg\max U(e). \]
 The principal's problem can be written as:
\[
\max_{\bar{t}, \underbar{t}} U_P(e) \]
subject to:
\[
U(e, \{\bar{t}, \underbar{t}\} ) \geq 0\]
and
\[
e = \arg\max U(e, \{\bar{t}, \underbar{t}\} ). \]

The first-order condition for the agent's problem is given by:
\[
\frac{dU}{de} = p'(e) (\bar{t} - \underbar{t}) - c'(e) = 0 \]

which we can rearrange to get:
\[
\bar{t} =\frac{ c_0}{p'(e)} + \underline{t}  \]

This can be substituted into the individual rationality constraint to get:
\[
\underline{t} \geq c(e) - \frac{p(e)}{p'(e)}c_0.  \] 
With limited liability, we also have that \(\underline{t} \geq 0\). This gives us the following contract:
\[\underline{t} = \max\left(0, c(e) - \frac{p(e)}{p'(e)}c_0\right),\, \bar{t} =  \max\left(\frac{ c_0}{p'(e)} , c(e) + \frac{1-p(e)}{p'(e)}c_0\right) \]

We assume that limited liability is binding, so that \(\underline{t} = 0\) and \(\bar{t} = \frac{ c'(e)}{p'(e)} \). 

Notice how this is similar to the case of the two-effort model, where the wage for a low outcome is given by:
\[
\underline{t} = \frac{\pi_0}{\pi_1 - \pi_0}c, \bar{t} = \frac{c}{\pi_1 - \pi_0} .\]

\section{Ongoing work and conclusions}

The results and framework presented in this paper offer a new lens through which  individual investments into improving generative AI systems might be viewed. While our model has been chosen to be as parsimonious as is possible  while still demonstrating the key economic tradeoffs, a hallmark of our framework is that it does not make assumptions about the nature of the AI implementation or work it supports. As generative AI continues to evolve beyond current large language models and towards more sophisticated multimodal systems and advanced reasoning capabilities, we believe our simple framework will remain relevant. 

We show that the tradeoffs associated with a strategic human’s incentives to actively contribute to the creation of better generative AI systems are real. Even when one's AI twin enables the production or higher-quality output or a reduction in effort, incentives to improve this twin are often constrained by the risk of replacement or the potential lowering of earnings. While counterintuitive at first glance, the finding --- that training an AI twin is incentive-aligned only when the stand-alone gains grow faster than those when the human and AI working in conjunction --- makes sense when one considers a worker's desire to preserve their wage bargaining power. 

Further, our finding that the strategic imperative a human might have to contribute less to the AI twin highlights the importance of careful incentive alignment when aiming to induce human workers to contribute to the improvement and customization of generative AI systems. As our two-period model shows, in the short run, workers may mistakenly embrace and overinvest in these training efforts, but once it becomes clear that those investments that yield the highest immediate personal productivity gains also put one at the greatest risk of mistakenly creating an AI twin that will displace them in the long run, this behavior will change. 

Our ongoing work is focused on extending and generalizing our model while also refining its underlying framework. Our core results generalize naturally to a setting involving continuous outcomes where more training leads to an outcome distribution that is better in the sense of first-order stochastic dominance. However, this extension also allows us to define the nature of the work and the technology in a more nuanced manner by expanding the ways in which we can differentiate the distribution of outcomes of the stand-alone AI twin from the human-AI combination, considering complementarity versus substitution and also allowing a richer specification of verifiability that separates the stand-alone AI twin's outcome distribution from that of the human-AI combination even when the human exerts low effort.  

Reinterpreted appropriately, our results could also suggest that effective AI tools will incorporate features that make outputs more transparent and assessable by non-experts, while also providing robust editing capabilities that allow for meaningful human intervention. This interpretation provides another argument in favor of a shift away from black-box solutions toward more interpretable and malleable AI systems. Strengthening this kind of conclusion from our analytical results will require an approach that unpacks the connection between the tasks comprising the work being done and the capabilities of the supporting AI, and models them separately. 

The empirical literature connecting generative AI to productivity changes is growing rapidly. As we encounter new results that deepen our understanding of the nature of interaction between generative AI and human workers, it is possible that other aspects of work beyond editability and verifiability could be relevant. 

As the returns from improvements to general-purpose generative AI systems are increasingly constrained by the plateauing of "scaling laws," by the unavailability of increasingly larger training data sets, and by the specter of "model collapse," engaging human workers in the improvement of AI systems will be increasingly important for organizations to continue to realize productivity gains from generative AI. The realization that the recent performance improvements of DeepSeek over OpenAI's systems stem from not just the use of more sophisticated reasoning but from a heavier reliance on RLHF only underscore this importance. 

Preserving and sustaining improvements in company-specific generative AI will likely require nuanced design of human incentives. The viability of contracting that align a human's returns from contributing to improvements in AI through, for example, fractional IP ownership, may be countered by the associated transaction costs and challenges in attribution, but this nevertheless represents a  direction for future inquiry. Policy interventions that better align an organization's incentives to balance worker returns with generative AI improvements could  yield more reliable training inputs and more sustained long-run productivity gains for a country. 

\bibliographystyle{plainnat}
\bibliography{research}

\appendix
\section{Proofs}

\subsection{Proof of Proposition \ref{prop:no-inv}}
\textsc{Proposition.}
If, for all \(v \in [0, v_{\max}]\), 
    \begin{equation}
        \frac{c'(v)}{c(v)} < \frac{Q'_\pi(v)}{Q_\pi(v)-1}
    \end{equation}
    then the agent's optimal training investment is $v=0$. 
\textsc{Proof.}
    If $U'(v) < 0$ for all \(v \in [0, v_{\max}]\), then $U(v)$ is maximized at \(v=0\). 
    Thus, we want to find conditions under which $U'(v) < 0$. Recall that we defined $U(v) = \frac{c(v)}{Q_\pi(v)-1}$. Therefore,
    \[U'(v) = \frac{c'(v) (Q_\pi(v)-1) - c(v) Q'_\pi(v)}{(Q_\pi(v)-1)^2}.\]
    Now, recall that $Q_\pi(v) = \frac{\pi_1(v)}{\pi_0(v)}$, and by assumption, $\pi_1(v) > \pi_0(v)$ for all \(v \in [0, v_{\max}]\). Thus we know that if \(Q_\pi(v) > 1\), and it follows that  \(Q_\pi(v)-1 > 0\).

    Thus, the denominator of $U'(v)$ is always strictly positive, and thus, the inequality $U'(v) < 0$ is equivalent to 
    \[c'(v) (Q_\pi(v)-1) - c(v) Q'_\pi(v) < 0.\]
    By assumption, $c(v) > 0$ for all $v$, so we can arrange the inequality to 
    \[\frac{c'(v)}{c(v)} < \frac{Q'_\pi(v)}{Q_\pi(v)-1}.\]

    The result follows.

\subsection{Proof of Proposition \ref{prop:max-inv}}
\textsc{Proposition.}    
If, for all \(v \in [0,v_{\max}]\), \(Q_\pi'(v) \leq 0\) and 
    \begin{equation}
        \bigg \lvert \frac{c'(v)}{c(v)} \bigg \rvert < \bigg \lvert \frac{Q_\pi'(v)}{Q_\pi(v)-1} \bigg \rvert
    \end{equation} 
    then the agent's optimal training investment is \(v=v_{\max}\).

\textsc{Proof.} 
If $U'(v) > 0$ for all \(v \in [0, v_{\max}]\), then $U(v)$ is maximized a $v=v_{\max}$. Thus, we want to find conditions in which $U'(v) > 0$ for all $v$. As above, we take the derivative of $U(v)$ with respect to $v$,

    \[U'(v) = \frac{c'(v) (Q_\pi(v)-1) - c(v) Q'_\pi(v)}{(Q_\pi(v)-1)^2}.\]

    Since the denominator of $U'(v)$ is strictly positive as established in the previous proof, the inequality $U'(v) > 0$ is equivalent to 
    \[c'(v) (Q_\pi(v)-1) - c(v) Q'_\pi(v) > 0.\]

    By assumption, $c(v) > 0$ for all $v$, so we can arrange the inequality to 
    \[\frac{c'(v)}{c(v)} > \frac{Q'_\pi(v)}{Q_\pi(v)-1}.\]

    We have also assumed that $c'(v) \leq 0$ and since $c(v) > 0$, we also know that $0 \geq \frac{c'(v)}{c(v)} >\frac{Q'_\pi(v)}{Q_\pi(v)-1} $. Thus $\frac{Q'_\pi(v)}{Q_\pi(v)-1}$ is negative, which is equivalent to $Q'_\pi(v) < 0$.

    Both sides of the inequality are negative, so we can rewrite it as 
    \[-\bigg \lvert \frac{c'(v)}{c(v)} \bigg \rvert > - \bigg \lvert \frac{Q_\pi'(v)}{Q_\pi(v)-1} \bigg \rvert\]
    and negate the entire inequality to obtain
    \[\bigg \lvert \frac{c'(v)}{c(v)} \bigg \rvert < \bigg \lvert \frac{Q_\pi'(v)}{Q_\pi(v)-1} \bigg \rvert.\]

    The result follows.

\subsection{Proposition \ref{prop:interior}}
\textsc{Proposition.}
A set of sufficient conditions for the agent to choose an interior value of \(v \in (0,v_{\max})\) are that:
    \begin{equation}
        \frac{c'(0)}{c(0)} > \frac{Q_\pi'(0)}{Q_\pi(0)-1},
     \end{equation} 
     and
     \begin{equation}
        \frac{c'(v_{\max})}{c(v_{\max})} < \frac{Q_\pi'(v_{\max})}{Q_\pi(v_{\max})-1}.
    \end{equation}

\textsc{Proof.}
To show an interior maximum exists, we first evaluate the derivative $U'(v)$ at the endpoints $v=0$ and $v=v_{\max}$.

At \( v = 0 \), using the proof from Proposition \ref{prop:max-inv}, the first condition \( \frac{c'(0)}{c(0)} > \frac{Q_\pi'(0)}{Q_\pi(0)-1} \) implies the numerator of \( U'(0) \), \( c'(0)(Q_\pi(0)-1) - c(0)Q_\pi'(0) \), is positive. Since the denominator \( (Q_\pi(0)-1)^2 > 0 \), it follows that \( U'(0) > 0 \). 

At \( v = v_{\max} \), the condition \( \frac{c'(v_{\max})}{c(v_{\max})} < \frac{Q_\pi'(v_{\max})}{Q_\pi(v_{\max})-1} \) implies the numerator of \( U'(v_{\max}) \), \( c'(v_{\max})(Q_\pi(v_{\max})-1) - c(v_{\max})Q_\pi'(v_{\max}) \), is negative. Thus, \( U'(v_{\max}) < 0 \).

Since $U'(0) > 0$, we know that $U(v)$ is increasing at 0, which means $v=0$ cannot be optimal. Similarly, since $U'(v_{max}) < 0$, we know $U(v)$ is decreasing at $v_{\max}$, which means $v=v_{\max}$ cannot be optimal.

We assume \( U'(v) \) is continuous on \( [0, v_{\max}] \). Then there exists some \(\hat{v} \in (0, v_{\max})\) where \( U'(\hat{v}) = 0 \). Furthermore, \( U'(v) \) transitions from positive to negative as \( v \) increases from \( 0 \) to \( v_{\max} \), which means \(\hat{v}\) is a local maximum. Therefore, $U'(\hat{v})=0$ must yield higher utility than either endpoint, and thus  the optimal solution is interior.

\subsection{Proof of Lemma \ref{lma:increasing-t}}
\textsc{Lemma.}    
If $\bar{t}'(v) > 0$, then $Q_\pi'(v) > 0$ and $U'(v) > 0$.

\textsc{Proof.} 
First, we show that $t'(v)>0$ implies $Q_\pi'(v)<0$:

Using the definition of $t(v) = \frac{c(v)}{\Delta\pi(v)}$, we obtain
\[t'(v)=\frac{c'(v)\Delta\pi(v)-c(v)\Delta\pi'(v)}{[\Delta\pi(v)]^2}.\]
If the above expression is positive, since $\Delta\pi(v)>0$, its numerator must also be positive:
\[c'(v)\Delta\pi(v)-c(v)\Delta\pi'(v)>0.\]
Since $c'(v)\le 0$, it follows that $\Delta\pi'(v)<0$, i.e., $\Delta\pi(v)$ is strictly decreasing.

Now, re-writing \(Q_\pi(v)=1+\frac{\Delta\pi(v)}{\pi_0(v)},\)
its derivative is
\[Q_\pi'(v)=\frac{\Delta\pi'(v)\pi_0(v)-\Delta\pi(v)\pi_0'(v)}{[\pi_0(v)]^2}.\]
Because $\Delta\pi'(v)<0$ and $\pi_0'(v) \ge 0$ is increasing, it follows that $Q_\pi'(v) < 0$.

Next, we can show that $U'(v)>0$ by rewriting \(U(v)=\pi_0(v)t(v)\), and differentiating to yield
\[U'(v)=\pi_0'(v)t(v)+\pi_0(v)t'(v).\]
Because $\pi_0(v)>0$, $t(v)>0$, $\pi_0'(v)\ge 0$, and by hypothesis $t'(v)>0$, both terms are nonnegative (with the second term strictly positive). Hence,
\[U'(v)>0.\]
The rest of the lemma follows from Proposition \ref{prop:max-inv}.

\end{document}